\documentclass[traditabstract,rnote]{aa}
\usepackage{graphicx}
\usepackage{txfonts}
 \newcommand{\Mo}{$~\mathrm{M_\odot}$}
 \newcommand{\Ro}{$~\mathrm{R_\odot}$}

\begin{document}
   \title{V456 Ophiuchi and V490 Cygni: \\ Systems with the shortest apsidal-motion periods}

   \titlerunning{V456 Oph and V490 Cyg: Apsidal motion systems}

   \author{P. Zasche \and
          M. Wolf}

   \authorrunning{P. Zasche \& M. Wolf}

   \institute{Astronomical Institute, Faculty of Mathematics and Physics, Charles University Prague,
   CZ-180 00 Praha 8, V Hole\v{s}ovi\v{c}k\'ach 2, Czech Republic, \email{zasche@sirrah.troja.mff.cuni.cz}\\}

   \date{\today}

% \abstract{}{}{}{}{}
% 5 {} token are mandatory

  \abstract{

Our main aim is the first detailed analysis of the two eclipsing binaries V456~Oph and V490~Cyg.
The system V456 Oph has been studied both photometrically via an analysis of its light curve
observed by the INTEGRAL/OMC and by the period analysis of all available times of minima. V490~Cyg
has been studied by means of a period analysis only. Many new times of minima for both systems have
recently been observed and derived. This allows us for the first time to study in detail the
processes that affect both binaries. The main result is the discovery that both systems have
eccentric orbits. For V456~Oph we deal with the eccentric eclipsing binary system with the shortest
orbital period known (about 1.016~day), while the apsidal motion period is about 23~years. V490~Cyg
represents the eclipsing system with the shortest apsidal motion period (about 18.8~years only).
The two components of V456~Oph are probably of spectral type F. We compare and discuss the V456~Oph
results from the light curve and the period analysis, but a more detailed spectroscopy is needed to
confirm the physical parameters of the components more precisely.}


   \maketitle
%
%________________________________________________________________

\section{Introduction}

The eccentric eclipsing binaries (EEBs) provide a great opportunity for studying the stellar
structure of the stars as well as testing the General Relativity outside the solar system. The
$O-C$~diagram analysis, which investigates the revolution of the line of apsides in the system has
been described elsewhere, e.g. \cite{1983Ap&SS..92..203G}, \cite{1995Ap&SS.226...99G}.
Nevertheless, new contributions to this topic with new systems are still welcome, especially for
cases where the apsidal motion period is adequately short and a few periods are covered. This is
the case for the two somewhat neglected systems \object{V456 Oph} and \object{V490 Cyg}.

\subsection{V456~Oph}

V456 Oph (= AN~108.1935 = SAO~123842, $V_{max} = 9.95$~mag) has been discovered as a variable star
by \cite{1935AN....255..401H}, with the remark that it is a "short-periodic one, but probably not
rapidly changing". After than \cite{1936AN....260..393G} incorrectly classified the star as a
$\delta$~Cep one with a preliminary period of about 14.6~d. No such variation has been detected
with the present data. The only spectral classification is that by \cite{1956ApJ...123..246R}, who
indicated the spectral type A5, but with a remark that because of underexposed plates and uncertain
ephemerides this classification is not very secure.

Although the first photoelectric light curve has been published by \cite{1988IBVS.3237....1D},
there was no light curve analysis of the system performed until today. The same applies to the
spectroscopic analysis, which has not yet been carried out, so the mass ratio of the pair in not
known. \cite{2006MNRAS.370.2013S} included the binary in the catalogue of systems located in the
instability strip, which means that is possibly contains a $\delta$~Scu component. However, no
indication of pulsations in V456~Oph has been detected.

\subsection{V490~Cyg}

V490 Cyg (= AN~76.1939, $V_{max} = 12.81$~mag) is an Algol-type eclipsing binary, even though the
SIMBAD database lists V490~Cyg as a $\beta$~Lyrae one. The system has been discovered as a variable
by \cite{Wachmann}, and its light curve coverage is too poor for any reliable analysis. Its
spectral type was derived to be F8, while \cite{Svechnikov} give an estimate of the spectral types
F8+[G4]. \cite{1988BICDS..35...15H} included this system in his list of stars with possible apsidal
motion, but since then it was not studied in detail. Other credible information about the physical
properties of the components is missing because there is so little spectroscopy and photometry of
this system.

\section{The period analysis}

\subsection{V456 Oph}

The set of published times of minima for V456~Oph is quite extensive, covering more than 70 years.
Regrettably, the old minima are only photographic and their scatter is so large that one cannot use
them for any reliable analysis. We used only the more precise photoelectric and CCD ones, which
were published after 1970. These minima roughly follow the linear ephemerides given in GCVS, but
there some variations are clearly visible.

We tried to collect all available minima times and also to derive some new ones. A few of the
already published ones were recalculated once again and corrected for the final analysis. Besides
these minima times, we also used the photometry of V456~Oph obtained with the robotized and
automated telescopes working today. These are

\begin{itemize}
   \item ASAS - the automated survey, $V$ filter, \cite{2002AcA....52..397P}, \texttt{http://www.astrouw.edu.pl/asas/}
   \item OMC - the OMC camera onboard the INTEGRAL satellite, using the $V$ filter, \cite{2004ESASP}, \texttt{https://sdc.laeff.inta.es/omc/}
   \item Pi of the sky - the automated telescope, unfiltered, \cite{2005NewA...10..409B}, \texttt{http://grb.fuw.edu.pl/pi/}
\end{itemize}

\begin{figure}
  \centering
  \includegraphics[width=91mm]{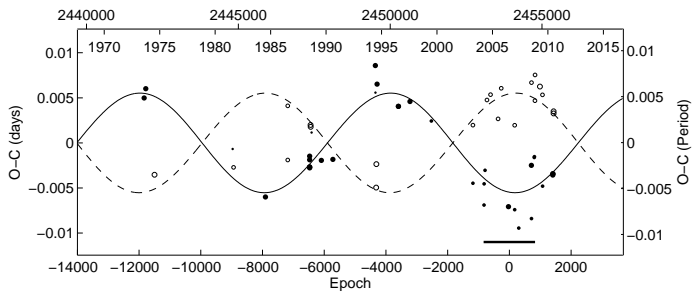}
  \caption{$O-C$ diagram of V456 Oph. The lines represent the fit according to the apsidal motion
  hypothesis (see text and Table \ref{OCparam1}), the solid line stands for the primary, while the dashed
  line stands for the secondary minima, dots stand for primary and open circles for the secondary minima.
  The black line near the bottom axis represents the time interval covered with the OMC data used for the
  light curve analysis.}
  \label{Fig3}
\end{figure}

Sixteen new minima times were derived from these surveys, and some of the published ones were
computed again (see Table \ref{Minima}). Our new minima times were observed in the Ond\v{r}ejov
observatory with the 65-cm telescope. We used the \cite{Kwee} method for all these minima. The mean
linear light elements suitable for observations are
\begin{equation}
  \mathrm{Prim.~Min.} = 2453923.9358 + 1.01600124 \cdot \mathrm{E},
\end{equation}
which were also used for deriving the proper epochs and types of the minima times written in Table
\ref{Minima}.

If we plot these data points in the $O-C$ diagram, the difference between primary and secondary
minima is clearly visible. Following the method of apsidal motion analysis as described in
\cite{1983Ap&SS..92..203G}, we tried the computation with orbital inclination $i=90^\circ$ for the
first time. After that we used for the second attempt the inclination $i=87.88^\circ$ (see Sect.
\ref{LCanal}), which resulted in almost the same parameters (owing to the term $\cot^2(i)$, which
is nearly 0 for our inclination). In Fig. \ref{Fig3} we plot the final fit with the apsidal motion
hypothesis on all used data points. This leads to the parameters of the motion of apsides given in
Table \ref{OCparam1}. All uncertainties of the parameters were calculated from the covariance
matrix of the fit and from the uncertainty of the inclination. Evidently the apsidal period is very
short, about only 23~yr, which places this system among one those with the shortest apsidal motion.

\begin{table}
 \caption{New and recalculated CCD minima times of V456 Oph.}
 \label{Minima}
 \scriptsize
 \centering
 \begin{tabular}{l l l l l}
 \hline\hline
  HJD - 2400000 & Error & Type & Filter & Observer/Reference\\ %$^{\mathrm{a}}$ \\
 \hline
 48113.4210   & 0.003   & Pri & C & A.Paschke, BBSAG 96 \\
 48113.42288  & 0.00108 & Pri & C & A.Paschke - recalculated \\
 52724.03387  & 0.0011  & Pri & V & ASAS \\
 52724.54831  & 0.0012  & Sec & V & ASAS \\
 53089.7944   & 0.0003  & Pri & V & P.Sobotka-OMC, IBVS 5809 \\
 53089.79189  & 0.0012  & Pri & V & OMC - recalculated \\
 53091.8267   & 0.0004  & Pri & V & P.Sobotka-OMC, IBVS 5809 \\
 53091.82625  & 0.00037 & Pri & V & OMC - recalculated \\
 53123.32379  & 0.00045 & Pri & V & OMC \\
 53188.86367  & 0.0044  & Sec & V & ASAS \\
 53305.70440  & 0.0042  & Sec & V & OMC \\
 53553.60602  & 0.0019  & Sec & V & ASAS \\
 53657.24154  & 0.00059 & Sec & V & OMC \\
 54099.69661  & 0.0034  & Pri & V & ASAS \\
 54099.19798  & 0.0028  & Sec & V & ASAS \\
 54232.79073  & 0.0009  & Pri & C & Pi of the sky\\
 54653.93939  & 0.0012  & Sec & V & ASAS \\
 54654.43230  & 0.0037  & Pri & V & ASAS \\
 54742.8314   & 0.0009  & Pri & V & P.Zasche-OMC, IBVS 5931 \\
 54742.83119  & 0.00045 & Pri & V & OMC - recalculated \\
 54749.94329  & 0.00032 & Pri & V & OMC \\
 54766.7134   & 0.0012  & Sec & V & P.Zasche-OMC, IBVS 5931 \\
 54766.71352  & 0.00032 & Sec & V & OMC - recalculated \\
 54769.7640   & 0.0011  & Sec & V & P.Zasche-OMC, IBVS 5931 \\
 54769.76438  & 0.00066 & Sec & V & OMC - recalculated \\
 54930.29129  & 0.00056 & Sec & V & OMC \\
 55016.65047  & 0.0018  & Sec & V & ASAS \\
 55017.14834  & 0.0035  & Pri & V & ASAS \\
 55352.42998  & 0.00003 & Pri & R & This paper \\
 55356.49413  & 0.00004 & Pri & R & This paper \\
 55357.51014  & 0.00008 & Pri & R & This paper \\
 55379.36085  & 0.00005 & Sec & R & This paper \\
 55382.40909  & 0.00019 & Sec & R & This paper \\
 \hline
\end{tabular}
\end{table}

\begin{table}[b]
 \caption{The parameters of the apsidal motion fit for V456~Oph and V490~Cyg.}
 \label{OCparam1}
 \small
 \centering
 \begin{tabular}{c c c c}
 \hline\hline
 Parameter & V456~Oph & V490~Cyg \\
 \hline
  $HJD_0$   & 2453923.9358$\,$(27) & 2451491.5931$\,$(51) \\
  $P$ [d]   & 1.01600124$\,$(24)   & 1.14023698$\,$(23) \\
  $P_a$ [d] & 1.01612627$\,$(25)   & 1.14042668$\,$(23) \\
  $e$       & 0.017$\,$(9)         & 0.045$\,$(15) \\
  $\omega$ [deg] & 351.1$\,$(1.6)  & 342.42$\,$(3.4) \\
  $\dot \omega$ [deg/cycle] & 0.044$\,$(3) & 0.060$\,$(12) \\
  $U$ [yr]  & 22.6$\,$(1.3)        & 18.8$\,$(3.2) \\
 \hline
 \end{tabular}
\end{table}

\subsection{V490 Cyg}

The system V490~Cyg has much lower published times of minima observations. The first rough
times-of-minima estimates are those by \cite{Wachmann2} from his photometry in the 1930's, but
these have such a large scatter that they cannot be used for any reliable analysis. However, he
also noticed that the secondary minimum is not symmetric with regards to the primary one.
Nevertheless, a possible eccentricity and apsidal motion have never been studied since then. The
more precise photoelectric and CCD observations have been measured since 1999, but there are only
12 published minima.

A few new CCD observations were obtained in the Ond\v{r}ejov observatory with the same telescope as
for V456~Oph, and two new minima times were also derived from the INTEGRAL/OMC data. The new
measurements and the already published ones are presented in Table \ref{Minima2}. The suitable
linear ephemerides for observations are
  \begin{eqnarray}
     \mathrm{Prim.~Min.} & = & 2451491.6075 + 1.14023698 \cdot \mathrm{E}, \\
     \mathrm{\,\,Sec.~Min.} & = & 2451491.5802 + 1.14023698 \cdot \mathrm{E}.
 \end{eqnarray}

\begin{table}[b]
 \caption{New and already published minima times of V490 Cyg.}
 \label{Minima2}
 \scriptsize
 \centering
 \begin{tabular}{l l l c l}
 \hline\hline
  HJD - 2400000 & Error & Type & Filter & Observer/Reference\\
 \hline
 51487.6184 & 0.0002  & Sec  & V   & D.B.Caton - IBVS 5595 \\
 51491.5776 & 0.0004  & Pri  & V   & D.B.Caton - IBVS 5745 \\
 51495.5994 & 0.0002  & Sec  & V   & D.B.Caton - IBVS 5745 \\
 52612.43506& 0.0047  & Pri  & V   & Integral/OMC \\
 52613.024  & 0.002   & Sec  & V   & P.Sobotka - IBVS 5809 \\
 52813.7080 & 0.0002  & Sec  & V   & D.B.Caton - IBVS 5595 \\
 52841.6237 & 0.0002  & Pri  & V   & D.B.Caton - IBVS 5595 \\
 53256.6755 & 0.0001  & Pri  & C   & T.Krajci - IBVS 5690 \\
 53260.6752 & 0.0003  & Sec  & C   & T.Krajci - IBVS 5690 \\
 53660.3244 & 0.0005  & Pri  & I   & F.Agerer - IBVS 5731 \\
 53660.331  & 0.0005  & Pri  & V   & R.Diethelm - IBVS 5713 \\
 53913.46115& 0.0002  & Pri  & R   & This paper \\
 53934.54462& 0.0010  & Sec  & R   & This paper \\
 54335.35387& 0.0013  & Pri  & R   & This paper \\
 54685.4130 & 0.0046  & Pri  & I   & F.Agerer - IBVS 5889 \\
 54694.53781& 0.0032  & Pri  & V   & Integral/OMC \\
 55096.4362 & 0.0004  & Sec  & I   & F.Agerer - IBVS 5941 \\
 55376.3966 & 0.0002  & Pri  & R   & This paper \\
 55392.35998& 0.0002  & Pri  & R   & This paper \\
 55405.44030& 0.0002  & Sec  & R   & This paper \\
 55445.34892& 0.0002  & Sec  & R   & This paper \\
 55462.45268& 0.0002  & Sec  & R   & This paper \\
 55470.43426& 0.0002  & Sec  & R   & This paper \\
  \hline
\end{tabular}
\end{table}

The minima times presented in Table \ref{Minima2} were used for the period analysis, which we did
by applying the apsidal motion hypothesis. The only difference in analysis between V490~Cyg and
V456~Oph was the assumption of an inclination $i=90^\circ$ for V490~Cyg because we had no light
curve analysis. The difference between primary and secondary is clearly visible, reaching up to 47
minutes, which is surprisingly high for a binary with such a short orbital period. The analysis led
to the parameters of the apsidal motion presented in Table \ref{OCparam1}. Obviously the resulting
value of the apsidal motion period of about 18.8~years is even shorter than for V456~Oph; we are
therefore dealing with the shortest apsidal motion period known among the EEBs today.

\begin{figure}
  \centering
  \includegraphics[width=91mm]{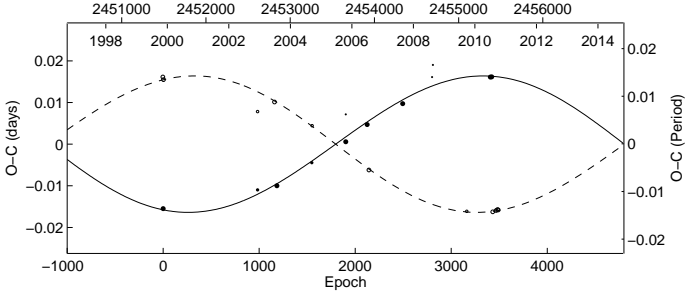}
  \caption{$O-C$ diagram of V490 Cyg. The lines represent the fit according to the apsidal motion
  hypothesis (see text and Table \ref{OCparam1}), the solid line stands for the primary, while the dashed
  line stands for the secondary minima, dots stand for the primary and open circles for the secondary minima.}
  \label{Fig4}
\end{figure}

\section{Light curve analysis} \label{LCanal}

The whole light curve of V456 Oph was observed with the OMC camera onboard the INTEGRAL satellite,
a description of which is given in \cite{2004ESASP}. The standard $V$ filter was used, but the
optical telescope has an aperture of only 5~cm in diameter. We obtained several hundred
observations, of which we used 449 for the analysis.

The programme {\sc PHOEBE} (ver. 0.29, Pr{\v s}a \& Zwitter 2005), based on the Wilson-Devinney
algorithm (Wilson \& Devinney 1971) was used for the analysis. The "detached binary" mode (in
Wilson \& Devinney mode 2) was used with several assumptions. First, the ephemerides ($HJD_0$ and
$P$) and the apsidal motion parameters ($e$, $\omega$, and $\dot \omega$) were adopted from the
period analysis, because the minima times cover a longer time span, therefore these quantities are
derived with higher precision. Secondly, the mass ratio $q$ and temperature of the primary
component $T_1$ were set and the other relevant parameters were adjusted for the best fit. We
changed the $q$ and $T_1$ values in the wide range of values to obtain the best fit according to
the RMS value and also the physical plausibility of the fit. This means during that the fitting
process we scanned the parameter space in $q$ ranging from 0.1 to 1.2 and in $T_1$ from 15400~K to
6500~K.

We fitted the other light curve parameters, which are the luminosities $L_1$ and $L_2$ in the $V$
filer, the temperature of the secondary $T_2$, the inclination $i$, the Kopal's modified potentials
$\Omega_1$ and $\Omega_2$, the synchronicity parameters $F_1$ and $F_2$, the third light $l_3$. The
limb-darkening coefficients were automatically interpolated by the {\sc PHOEBE} programme from
van~Hamme's tables (see van Hamme 1993), using the linear cosine law for the values of $T_{eff}$
and $\log g$ of both components resulting from the analysis. The values of the gravity brightening
and bolometric albedo coefficients were set at their suggested values for convective atmospheres
(see Lucy 1968), i.e. $G_1 = G_2 = 0.32$, $A_1 = A_2 = 0.5$.

\begin{figure}
  \centering
  \includegraphics[width=87mm]{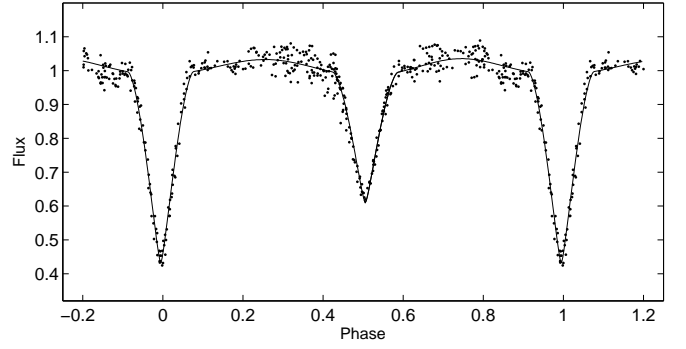}
  \caption{{\sc PHOEBE} light curve solution of V456 Oph based on the OMC data, the solid line
  represents our final solution (see the text and Table \ref{LCparam}).}
  \label{Fig1}
\end{figure}

\begin{table}[b]
 \caption{The light curve parameters of V456 Oph.}
 \label{LCparam}
% \scriptsize
 \small
 \centering
 \begin{tabular}{c c | c c}
 \hline\hline
 Parameter & Value & Parameter & Value \\
 \hline
  $T_1$ [K]  & 6840                 & $L_1 / (L_1+L_2)$ (V) & 59 $\pm$ 4 \% \\
  $T_2$ [K]  & 6700 $\pm$ 430       & $L_2 / (L_1+L_2)$ (V) & 41 $\pm$ 3 \% \\
  $q$ ($={M_2}/{M_1}$) & 0.96 $\pm$ 0.15 & $F_1$ & 1.097 $\pm$ 0.402 \\
  $i$  [deg] & 87.88 $\pm$ 0.81     & $F_2$ & 0.843 $\pm$ 0.356 \\
  $\Omega_1$ & 5.09 $\pm$ 0.32      & $x_1$ & 0.505 $\pm$ 0.021 \\
  $\Omega_2$ & 5.08 $\pm$ 0.29      & $x_2$ & 0.504 $\pm$ 0.019 \\ \hline
  $R_1/a$    & 0.249 $\pm$ 0.05     & $R_2/a$ & 0.247 $\pm$ 0.04 \\
 \hline
 \end{tabular}
\end{table}

The best fit was achieved with the light curve parameters given in Table \ref{LCparam}, and the
figure with the final fit is plotted in Fig.\ref{Fig1}. Nonzero eccentricity is clearly visible
from this plot, which is quite surprising in a binary with such a short orbital period. No other
EEB with a shorter period is known today. The value of the third light is \textbf{$l_3 = (0 \pm
4)~\%$}, which indicates that there is no other visible companion to the system in the $V$ filter
(under the assumption that this component is also located on the main sequence). We made several
attempts with nonzero values of the third light, but these did not lead to a satisfactory solution.

Because there is no spectroscopic analysis, the precise physical parameters cannot be computed
directly, but need to be roughly estimated with the assumption that both components are located on
the main sequence. We derived the following values: $M_1=1.46$\Mo, $M_2=1.41$\Mo, $R_1=1.51$\Ro,
$R_2=1.49$\Ro. These are only very preliminary values, but lead to spectral types of about F1 + F2
for the two components. We obtained roughly the same result (F0+F1) with the standard
mass-luminosity relation for the main sequence stars (e.g. Malkov 2007), applying the luminosity
ratio derived from the light curve analysis.

The parameters are very different from what one could expect for a main sequence star of spectral
type A5 (Roman 1956), and the masses are also different from those estimated by Brancewicz \&
Dworak (1980), but the presented values provided the best light curve fit, and the parameters of
the apsidal motion also agree well with the theoretical values (see below). The spectral type
presented by Roman (1956) was only estimated on the basis of poor photographic spectra. On the
other hand, there is also the $BVR$ photometry in the NOMAD catalogue (Zacharias et al. 2004), from
which $B-V = 0.279~$mag and $V-R = 0.165~$mag. These values indicate (Houdashelt et al. 2000) that
the temperature of the system is about 6500~K, therefore of the spectral type about of F5.

\section{Discussion}

For the light curve analysis of the system V456~Oph the ephemerides and the apsidal motion
parameters were fixed, but another approach could be to compute these parameters directly also from
the light curve. The problem is that the data coverage for the light curve is rather fairly in time
(about only 1/5 of the apsidal period), and the data for the light curve have relatively high
scatter as well.

The eclipsing binaries, V456~Oph and V490~Cyg, with their respective apsidal motion periods of
about only 20~years place these systems among a few unique ones with apsidal periods below 30~years
(see Table \ref{shortest}).

\begin{table}[b]
 \caption{The EEBs with the shortest apsidal motion period.}
 \label{shortest}
 \scriptsize
 \centering
 \begin{tabular}{c c c l c c}
 \hline\hline
  System     &  Spectr.  &    P [d]  & \multicolumn{1}{c}{e} &  U [yr] &  Reference \\
 \hline
  V490 Cyg   &           &    1.1402  & 0.045   &  18.8    &  This paper      \\
  V381 Cas   &     B3    &    1.7459  & 0.0253  &  19.74   &  Wolf et al. (2010)\\
  U Oph$^{\mathrm{*}}$&B5+B5& 1.6773  & 0.00305 &  20.88   &  Vaz et al.  (2007) \\
  V456 Oph   &   ?F1+F2? &    1.0160  & 0.017   &  22.6    &  This paper      \\
  GL Car$^{\mathrm{*}}$&B0+B1& 2.4222 & 0.146   &  25.2    &  Wolf et al. (2008) \\
  V478 Cyg   &   B0+B0   &    2.8809  & 0.0158  &  27.1    &  Wolf et al. (2006) \\
 \hline
 \end{tabular}
 \begin{list}{}{}
  \item [$^{\mathrm{*}}$] Triple system \\
 \end{list}
\end{table}

Our next task was to derive the averaged internal structure constant as well and to compare it with
the theoretical value. This task was done after subtraction of the relativistic term, which
resulted for V456~Oph in the value $\dot \omega_{\mathrm{rel}} = 0.0011~\mathrm{deg/cycle}$, about
only 2.5\% of the total apsidal motion rate. Therefore, the internal structure constant is
\vspace{-2pt} $$ \log k_{\mathrm{2,obs}} = -2.44 \pm 0.20. \vspace{-1pt} $$ The surprisingly high
value of the uncertainty is mainly caused by the error of the relative radii from the light curve
analysis. We can compare this value with the stellar evolution grids (e.g. by Claret 2004) and the
theoretical values of $k_{\mathrm{2,theor}}$. Using the value of $\log M = 0.1725$
($M=1.49~\mathrm{M_\odot}$), we obtained the value of \vspace{-2pt} $$ \log k_{\mathrm{2,theor}} =
-2.41 \pm 0.05 \vspace{-1pt} $$ for the main sequence star with an age between $0$ and
$1.5~\cdot~10^9$~yr. This could be interpreted as a rough estimation of maximum age for this
system. No other eccentric eclipsing binary with such a late spectral type is known today.
Therefore a detailed analysis of its spectra would be very welcome.

\section{Conclusions}

We performed the first detailed photometric and period analysis of the two eclipsing systems
V456~Oph and V490~Cyg, which yielded the parameters of the apsidal motion with periods of about
only 23 and 19~years. With the orbital period of V456~Oph of only about 1.016~days we are dealing
with the shortest orbital period among the apsidal motion systems, while the period of apsidal
motion of 18.8~years of V490~Cyg makes this system the shortest among the EEBs. However, because we
lack a spectroscopic analysis, some of the physical parameters were only roughly estimated and
apparently contradict each other. New times of minima observations as well as a detailed
spectroscopic analysis are needed.

\begin{acknowledgements}

Based on data from the OMC Archive at LAEFF, pre-processed by ISDC. We thank the "ASAS" team and
also the "Pi of the sky" team for making all of the observations easily public available. This work
was supported by the Czech Science Foundation grant no. P209/10/0715 and also by the Research
Programme MSM0021620860 of the Czech Ministry of Education. Mr. Anton Paschke is also acknowledged
for sending us his photometric data and also Mr. Kamil Hornoch for the observational assistance.
This research has made use of the SIMBAD database, operated at CDS, Strasbourg, France, and of
NASA's Astrophysics Data System Bibliographic Services.

\end{acknowledgements}

\end{document}